\documentclass[aps,prl,twocolumn,showpacs]{revtex4}

\usepackage[centertags]{amsmath}
\usepackage{amssymb}
\usepackage{graphicx}
 
\newcommand{\kk}{{\mathbf k}}
\newcommand{\kko}{{\mathbf k}_1}
\newcommand{\kkt}{{\mathbf k}_2}
\newcommand{\kkmo}{{\mathbf k}_{\text{m}1}}
\newcommand{\kkmt}{{\mathbf k}_{\text{m}2}}
\newcommand{\cmo}{c_{\text{m}1}}
\newcommand{\cmt}{c_{\text{m}2}}
\newcommand{\cmot}{c_{\text{m}1,\text{m}2}}
\newcommand{\nmo}{n_{\text{m}1}}
\newcommand{\nmt}{n_{\text{m}2}}
\newcommand{\nmot}{n_{\text{m}1,\text{m},2}}
\newcommand{\Tif}{{T_{if}}}
\newcommand{\vk}{\hat{k}}
\newcommand{\derfrac}[2]{\genfrac{}{}{}{}{d#1}{d#2}}
\newcommand{\derfract}[2]{\genfrac{}{}{}{}{d^3#1}{d#2}}
\newcommand{\TR}{T_R}
\newcommand{\Rc}{R_c}

\begin{document}

\title{Atom-Molecule Laser Fed by Stimulated Three-Body Recombination}

\author{Bogdan Borca}

\altaffiliation[On leave from the ]{Institute for Space Sciences,
Bucharest-Magurele 76900, Romania.}

\author{Josh W. Dunn}

\author{Viatcheslav Kokoouline}

\author{Chris H. Greene}

\affiliation{JILA and Department of Physics, University of Colorado,
Boulder, Colorado 80309-0440}

\date{Submitted May 14, 2003}

\begin{abstract}
Using three-body recombination as the underlying process, we propose a
method of coherently driving an atomic Bose-Einstein condensate (BEC)
into a molecular BEC.  Superradiant-like stimulation favors
atom-to-molecule transitions when two atomic BECs collide at a
resonant kinetic energy, the result being two molecular BEC clouds
moving with well defined velocities.  Potential applications include
the construction of a molecule laser.
\end{abstract}


\maketitle

One of the major goals of the recent research in BEC physics is the
creation of a molecular condensate~\cite{zoller2002}.  Many
applications are envisioned for such a system, including coherent
quantum chemistry~\cite{moore2002}.  One approach to forming a
molecular BEC utilizes an atomic condensate along with a method of
coherently driving pairs of atoms into (bound) molecular states.  Two
mechanisms have been proposed for this: (i) photoassociation, in which
an external laser couples atomic states to molecular
states~\cite{juli98}, and (ii) control of the scattering length via
external magnetic field ramps tuned near a Feshbach
resonance~\cite{timm99}.  Another problem of interest to the BEC
community is the realization of an atom laser~\cite{hagley99}.  A
proposed technique~\cite{kozuma99} for this is the application of a
time-dependent light pulse resulting in Bragg scattering of the
condensate cloud, in which a fraction of the condensate receives a
large momentum kick and attains velocity relative to the parent cloud.

In this Letter we propose a new mechanism for the formation of a
molecular BEC from an atomic BEC. This mechanism relies on a twofold
bosonic stimulation of the three-body recombination process --- both
by the number of atoms present and by the number of molecules.  We
show that two atomic condensates with a well-defined (resonant)
relative velocity will mutually stimulate transitions into molecular
states, resulting in the production of two coherent molecular clouds,
each with a well-defined momentum.

Consider first the elementary process involving only three isolated
atoms.  The atoms, initially in an unbound state, are converted
through recombination into a bound molecular state with $1/3$ of the
recoil energy, while a third atom carries away the remaining $2/3$.
Conservation of momentum requires that the resulting atom and molecule
have momenta of equal magnitude ($\hbar k_0$), but in opposite
directions.  The value of $k_0$ is fixed by conservation of energy to
be $k_0=\sqrt{4mE_0/3 \hbar^2}$, where $E_0$ is the molecular binding
energy and $m$ is the atomic mass.  This process is characterized by a
transition matrix element $\Tif$ between the initial and final
three-body states.  The spontaneous decay of an ultracold atomic
sample of density $n$ through three-body recombination is usually
characterized by a loss rate coefficient $K_3$, such that $dn/dt = -
3K_3 n^3/6$.  Here $n^3/6$ is the number of atomic triplets available
for recombination in the considered volume and the factor of 3 denotes
the atoms lost in each recombination.  An additional factor of 1/6
multiplies the right-hand side if all three atoms share the same
state, i.e., if we consider the decay of a pure
condensate~\cite{kagan85}. The connection between $\Tif$ and $K_3$ is
given by~\cite{fed96}
\begin{align}
  K_3 =
  \frac{6/\hbar}{(2\pi)^2 }\int
  \derfract{\kk}{E}
  \left| \Tif \right|^2\,
  \delta\!\!\left(E - \frac{3\hbar^2k_0^2}{4m}\right) dE
  =\frac{6 k_0  m}{\hbar^3}\,|\Tif|^2\,.
\nonumber
\end{align}
If the recombination forms a weakly bound level with $E_0 \approx
\hbar^2/(m a^2)$ then $k_0=2/(\sqrt{3} a)$, where $a$ is the $s$-wave
scattering length. In this case, working in the ultracold limit and
using momentum normalized states, Ref.[8] obtains $\Tif=\zeta
\,\hbar^2\, a^{5/2}/m$~\cite{fed96}, where $\zeta$ is a numerical
constant.  More generally $K_3$ has been shown to scale with the
fourth power of the scattering length in most cases,
$K_3=\beta(\hbar/m)\,a^4$~\cite{esry99,nielsen99} which for the weakly
bound level gives $\beta=(12 \zeta^2 )/(\pi \sqrt{3})$.
Ref.~\cite{fed96} finds $\beta\approx 23$, while a more rigorous
calculation, valid for a wide range of interatomic
potentials~\cite{esry99}, yields $\beta \approx 180$ for $a>0$ and
$\beta \approx 1014$ for $a<0$.  (We omit a $\sin ^2$ modulation
factor discussed in [9,10].)

\begin{figure}
  \includegraphics[width=2.4in]{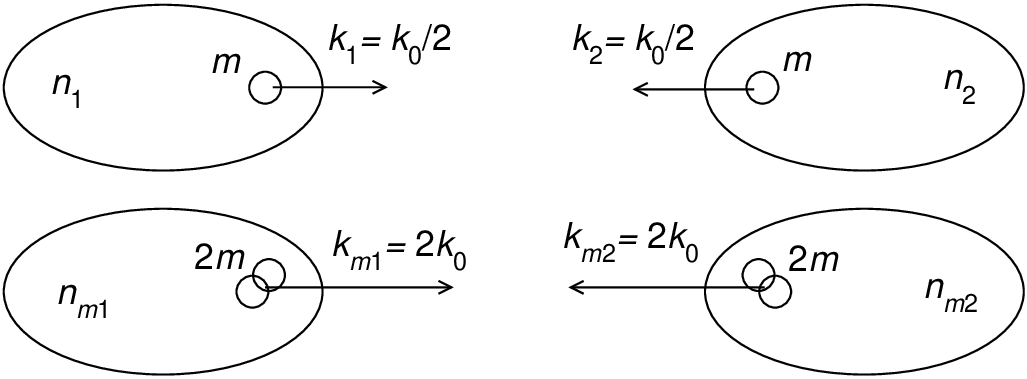}
  \caption{
    Schematic diagram of the four possible states of the atoms: atomic
    states with momenta $\hbar \kko$ and $\hbar \kkt$ have populations
    $n_1$ and $n_2$; molecular (molecular mass $2m$) states with momenta
    $\hbar \kkmo$ and $\hbar \kkmt$ have populations $\nmo$ and $\nmt$.
    }
  \label{popdiag}
\end{figure}

We are interested in the process in which three-body recombination is
stimulated by bosonic enhancement.  We consider two atomic condensate
clouds moving with relative velocity $\hbar k_0/m$, and choose a
reference frame in which the two clouds move with equal velocities
towards each other (see Fig.~\ref{popdiag}) along the direction $\vk$.
Let the wavevector of the atoms moving to the right (left) be denoted
by $\kko$ ($\kkt$) and the population of the mode by $n_1$ ($n_2$).
Note that $\kko=-\kkt=k_0/2\,\vk$.  Recombination of three atoms with
momentum $\hbar \kko$ can lead to a final state in which the resulting
atom has momentum $\hbar \kkt$, in which case the resulting molecule
has the momentum $\hbar \kk_{\text{m1}}=2\,\hbar k_0\,\vk$.  This
transition will be strongly favored due to bosonic stimulation if the
mode $\kkt$ is highly populated.  We recall the fact that the
magnitude of the momentum is fixed by energy and momentum conservation
and only its direction has an arbitrary value.  Therefore, it is
possible that the atom-molecule pair is ejected along a direction
different from $\vk$.  However, this process does not benefit from
bosonic stimulation.  By symmetry, recombination of three atoms with
momentum $\hbar \kkt$ will be strongly enhanced if the resulting
atom-molecule pair is ejected along the direction $\vk$. In this case
the momentum of the molecule has the value $\hbar \kkmt=-2\,\hbar
k_0\,\vk$ while the resulting atom has momentum $\hbar \kko$.  We
denote the numbers of molecules with wavevectors $\kkmo$ and $\kkmt$
by $\nmo$ and $\nmt$, respectively.  Finally, note that, for any
three-body recombination the reverse process is also possible.  For
example, an atom with momentum $\hbar \kk_{2}$ that collides with a
molecule with momentum $\hbar \kkmo$ can induce it to break up,
resulting in three atoms with momentum $\hbar \kko$.  In summary, the
four modes considered here are coupled by the transitions
\begin{equation}
  \begin{split}
    \label{trans}
    3\, \text{A}(\kko)\, &\rightleftharpoons \, \text{A}(\kkt)\, 
    + \, \text{A}_2(\kkmo), \\
    3\, \text{A}(\kkt)\, &\rightleftharpoons \, \text{A}(\kko)\, 
    + \, \text{A}_2(\kkmt).
  \end{split}
\end{equation}

The many-body interaction Hamiltonian describing the reactions in
Eqs.~(\ref{trans}) is
\begin{equation}
  \hat{H}_i = \hbar \omega\,
  \left(
    c_{2}^\dagger\,\cmo^\dagger\,c_1\,c_1\,c_1
    + c_{1}^\dagger\,\cmt^\dagger\,c_2\,c_2\,c_2
  \right)
  + \text{h.c.}
  \label{hi}
\end{equation}
Here $c_{1,2}$ ($\cmot$) are the annihilation operators for the two
atomic (molecular) modes and the coupling frequency is given by $\hbar
\omega =\Tif /V^{3/2}$, where $V$ is the quantization volume. For a
weakly bound molecule $\omega=[\zeta \hbar/(m a^2)] (a^3/V)^{3/2}$.
The two operators,
\begin{gather}
  \hat{N}
  =
  c_1^\dagger\,c_1\,
  +\,c_2^\dagger\,c_2\,
  +2\,\cmo^\dagger\,\cmo\,
  +2\,\cmt^\dagger\,\cmt\,,
  \nonumber
  \\
  \frac{\hat{P}}{(\hbar k_0/2)}
  = 
  c_1^\dagger\,c_1\,
  -\,c_2^\dagger\,c_2\,
  +4\,\cmo^\dagger\,\cmo\,
  -4\,\cmt^\dagger\,\cmt\,,
  \nonumber
\end{gather}
commute with the Hamiltonian in Eq.~(\ref{hi}) as a consequence of the
conservation of the total number of atoms and of the total momentum.

We now derive the equations of motion for the creation and
annihilation operators in the Heisenberg representation.  Adopting the
notation $a_{1,2}(\tau)$ and $b_{1,2}(\tau)$ for the atomic and
molecular Heisenberg operators divided by $\sqrt{N}$ (in which fast
oscillations with the frequency associated to the non-interacting
energy were factored out), the equations of motion read,
\begin{align}
  \label{opeq1}
  i\derfrac{}{\tau} a_1
  &=
  3\,a_1^\dagger\,a_1^\dagger\,b_1\,a_2\,+\,b_2^\dagger\,a_2\,a_2\,a_2 \,,
  \\
  \label{opeq2}
  i\derfrac{}{\tau} a_2
  &=
  3\,a_2^\dagger\,a_2^\dagger\,b_2\,a_1\,+\,b_1^\dagger\,a_1\,a_1\,a_1 \,,
  \\
  \label{opeq3}
  i\derfrac{}{\tau} b_1
  &=
  \,a_2^\dagger\,a_1\,a_1\,a_1 \,,
  \\
  \label{opeq4}
  i\derfrac{}{\tau} b_2
  &=
  \,a_1^\dagger\,a_2\,a_2\,a_2.
\end{align}
We use a scaled time $\tau=\Omega\,t$ with $\Omega=N^{3/2}\, \omega =
n^{3/2}\, \Tif$ ($n=N/V$ is the total concentration of atoms).  For a
shallow molecular level $\Omega= [\zeta \hbar/(ma^2)] (n a^3 )^{3/2}$.
Equations~(\ref{opeq1})--(\ref{opeq4}) describe the dynamic population
transfer between the atomic and molecular modes due to resonant
three-body recombination.  They are based on the assumption that all
two-body elastic collisions (atom-atom, molecule-molecule and
atom-molecule) and also collisional loss of the weakly bound molecules
to deeper molecular states can be neglected. In addition, we have
neglected the spatial and temporal changes of the slowly varying
envelopes of the condensate clouds.  The simplest approximation to the
above equation consists of replacing the operators by complex numbers
equal to their averages: $\tilde{a}_{1,2}(t) = \langle a_{1,2}(t)
\rangle$ and $\tilde{b}_{1,2}(t) = \langle b_{1,2}(t) \rangle$, i.e.,
the classical field approximation.
The resulting system of ordinary differential equations, with initial
conditions, can be integrated numerically.  We consider the
experimentally relevant case of an initial state with zero molecules.
We also assume that initially the $N$ atoms are evenly split between
the two momentum eigenstates: $ N_{1,2} \equiv n_{1,2}(t=0)=N/2$.
Assuming real initial conditions, we set $\tilde{ b}_{1,2}(0)=0$ and
$\tilde{a}_{1,2}(0)=1/\sqrt{2}$.  The approximate version of the
system of Eqs.~(\ref{opeq1})--(\ref{opeq4}) can be solved to describe
the population transfer from the atomic to the molecular modes in a
process characterized by the time constant $\TR= 2
\pi/\Omega$.  A characteristic that is similar to the superradiance
phenomenon is double amplification (stimulation): after an initial
buildup period for the molecular population, rates of further
transitions are proportional both to the number of existing molecules
in the recombined momentum state and to the number of atoms in their
relevant momentum state.

\begin{figure}
  \includegraphics[width=2.5in]{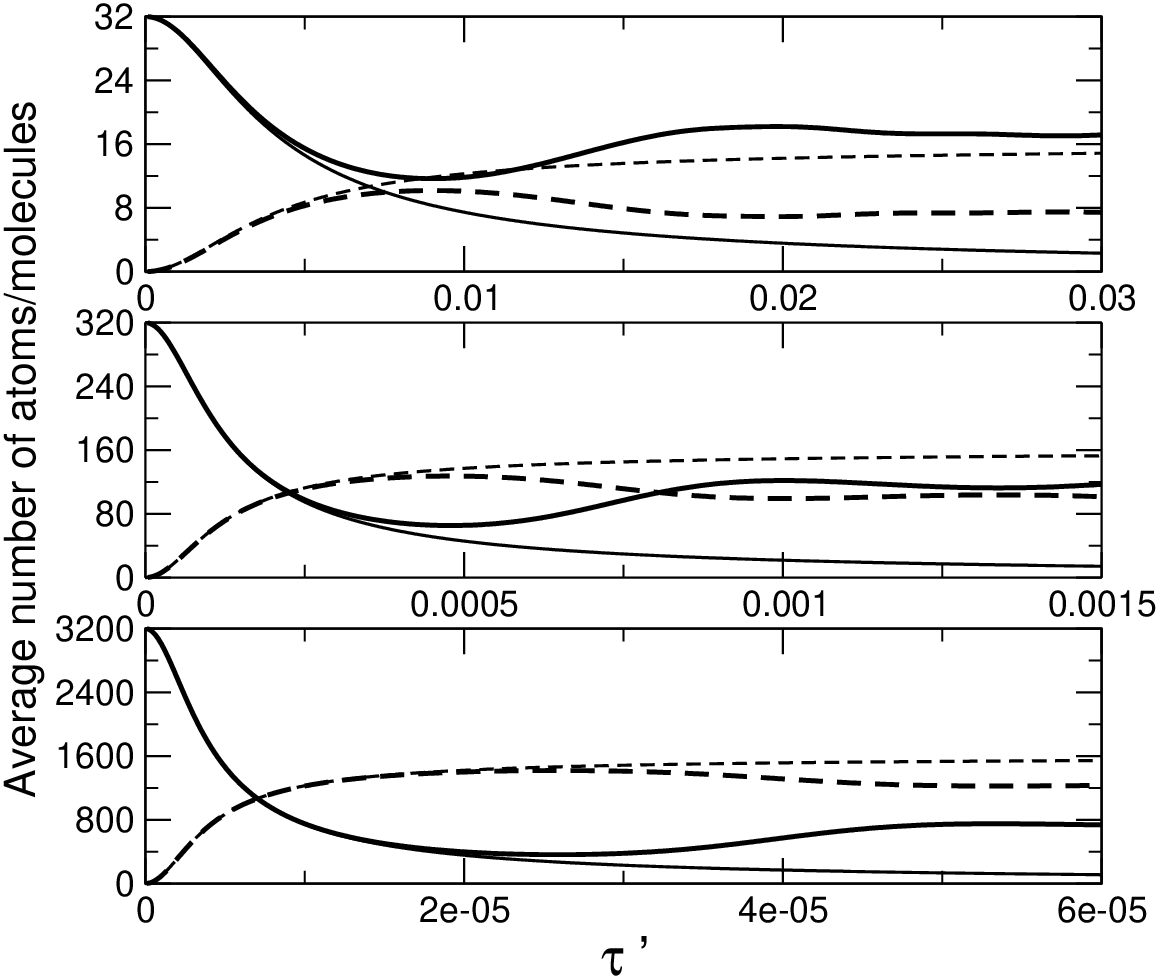}
  \caption{
    Comparison of the scaled-time-dependent average
    number of atoms (solid lines) and molecules (dotted lines),
    obtained in two ways: by solution of the approximate Heisenberg
    equations (thick lines) and by solution of the many-body
    Schr\"odinger equation (thin lines). The three panels correspond
    to different total numbers of atoms $N$: top $N=64$, middle
    $N=640$, bottom $N=6400$.
    }
\label{figcomp}
\end{figure}

The question of the validity of using the classical field
approximation can be studied in general by accounting for higher-order
terms in a cluster expansion of the four-operator products in
Eqs.~(\ref{opeq1})--(\ref{opeq4}) (see, e.g., Ref.~\cite{kohler02}).
However it is sufficient for our purpose to compare the results of the
approximate version of Eqs.~(\ref{opeq1})--(\ref{opeq4}) to the
many-body Schr\"odinger equation written using a basis of Fock states.
We designate these basis states by $|n_1,n_2,\nmo,\nmt\rangle$ where
$n_{1,2}$ ($\nmot$) are the occupation numbers of the two atomic
(molecular) modes.  Due to the two conservation laws $n_1 = N_1 + \nmt
- 3\nmo$ and $n_2 = N_2 + \nmo - 3\nmt$, we can restrict the basis set
to only those accessible by three-body recombinations from our initial
state and thus use only two occupation numbers, say $\nmo$ and $\nmt$,
to parameterize the Fock states.  With this notation, the many-body
wavefunction is
\begin{equation}
  |\Psi(t)\rangle=\sum_{\nmo,\nmt}
  f_{\nmo,\nmt}(t)\,|\nmo,\nmt\rangle.
\end{equation}
After eliminating the fast oscillations and introducing the time
scaling $\tau'=\omega\,t$, the Schr\"odinger equation becomes
\begin{align}
  \label{csch}
  &i \frac{d}{d\tau'} f_{\nmo,\nmt} 
  \\
  &= \sqrt{(n_1+1)(n_1+2)(n_1+3)n_2 \nmo}
  \;\;f_{\nmo-1,\nmt}
  \nonumber \\
  &+ \sqrt{n_1 (n_1-1)(n_1-2)(n_2+1)(\nmo+1) }
  \;\;f_{\nmo+1,\nmt}
  \nonumber \\
  &+ \sqrt{(n_2+1)(n_2+2)(n_2+3)n_1 \nmt}
  \;\;f_{\nmo,\nmt-1}
  \nonumber \\
  &+ \sqrt{n_2 (n_2-1)(n_2-2)(n_1+1)(\nmt+1) }
  \;\;f_{\nmo,\nmt+1}.
  \nonumber
\end{align}
Solving these equations with the initial conditions $f_{0,0}(0)=1$ and
$f_{\nmo,\nmt}(0)=0$ for $\nmo \neq 0$ and $\nmt \neq 0$, allows us to
calculate the average numbers of atoms populating any of the four
considered modes.
These averages are shown in Fig.~\ref{figcomp} for different values of
$N_1=N_2=N/2$. Comparison with the corresponding results obtained
using Eqs.~(\ref{opeq1})--(\ref{opeq4}) shows agreement over the
initial period of fast transitions of atoms into molecular states.  As
expected, the classical field approximation fails at shorter times for
lower values of $N$, while for sufficiently large $N$ it can be used
to accurately describe the initial transition burst converting atoms
to molecules.  Equation~(\ref{csch}) allows us to describe the
population transfer within the four-mode approximation without making
any additional approximations.  However, calculations using the Fock
basis become prohibitive if more than a few thousand atoms are
considered.

Experimental arrangements that use the process described above to
produce coherent molecular clouds can be readily envisioned.  The
simplest experiment would begin with a time-dependent Bragg splitting
of a single atomic cloud as in the experiment of Kozuma \emph{et
al.}~\cite{kozuma99}, but with the particular choice of momentum kick
that matches the three-body resonance condition.  In addition, the
value of the molecular binding energy can be tuned using an external
magnetic field~\cite{inouye98,Donley02}.  In order to take full
advantage of the stimulated transition into molecular states, (i.e.,
to transform a large fraction of the atomic clouds into molecular
clouds) the collision time of the two atomic clouds should be
comparable to the characteristic time of the three-body recombination
process $t_{\text{coll}} \sim \TR$.  Here
$t_{\text{coll}}\approx \Rc/(\hbar k_0/m)$ and $\Rc$
is the characteristic size of the condensate along the collision
direction.  Consider a sample of $^{85}$Rb with a concentration of
$10^{15}/{\rm cm}^3$.  Assuming that the scattering length is adjusted
with the help of a Feshbach resonance to $a=400$\,a.u. and that $E_0
=\hbar^2/m a^2$, then we estimate $\TR=0.45$\,ms (using $K_3$
given by~\cite{esry99}).  The resonant process described here is
characterized by a finite width if additional processes that limit the
lifetime of the involved states are considered.  One such process is
the collsionally induced decay of the molecular level to lower
molecular states.  Only a very rough estimation of the lifetime is
available for molecules formed by atomic species commonly used in BEC
experiments.  Some estimations (see, e.g,~\cite{timm99}) cite values
of at least 1\,ms for this lifetime at molecular densities of
$10^{15}/{\rm cm}^3$. In addition, recent experiments in
$^{85}$Rb~\cite{Donley02} suggest that this is a conservative
estimation, and that the actual lifetime might be somehow longer.

\begin{figure}
  \includegraphics[width=2.3in]{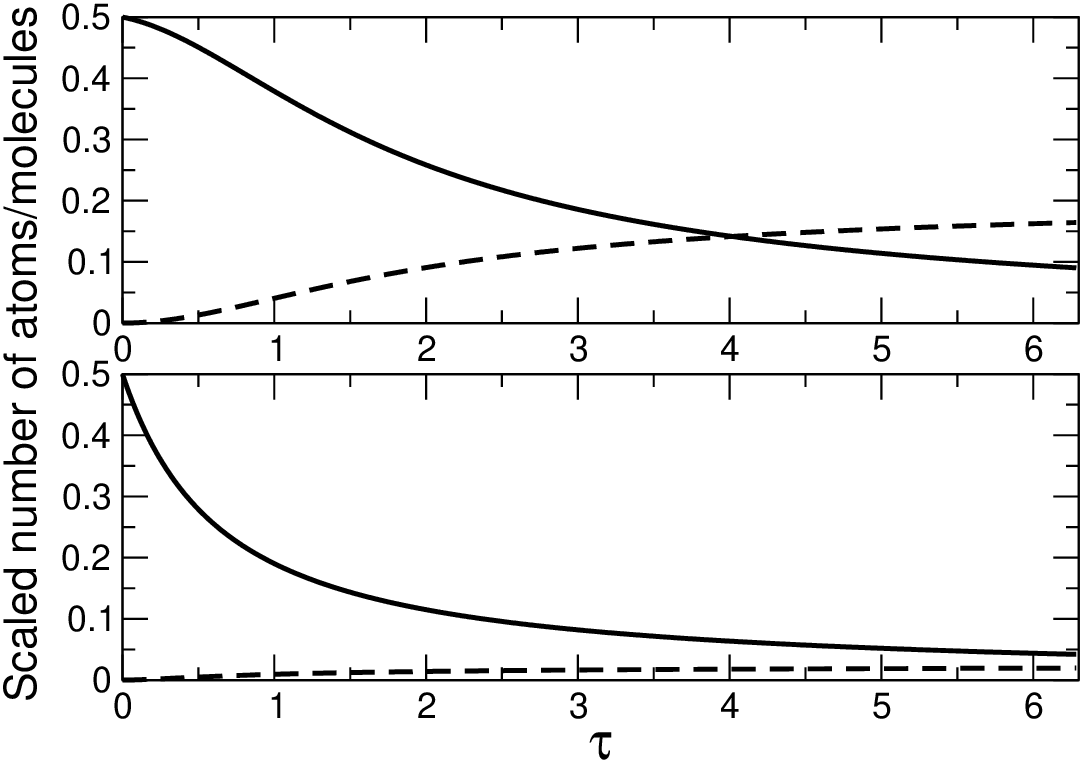}
  \caption{
    Solutions of Heisenberg equations with elastic loss (top panel:
    $\gamma=0.1$, bottom panel $\gamma=1.5$).  Solid line: atomic
    population fraction $|\tilde{a}_{1,2}(\tau)|^2$, dotted line:
    molecular population fraction $|\tilde{b}_{1,2}(\tau)|^2$.
    }
  \label{fighes}
\end{figure}

Up to this point we have neglected elastic collisions between atoms
and molecules.  In fact, if the resonant value of the relative
velocity is not enforced by the experimental setup, depletion of the
atomic clouds by elastic atom-atom scattering is the main loss process
taking place before the two condensate clouds
separate~\cite{band2000}.  As discussed in Ref.~\cite{band2000},
elastic scattering must be accounted for beyond the mean-field (i.e.,
Gross-Pitaevskii) approximation in order to observe the condensate
depletion due to elastic scattering.  We can include this process in
our simple model as a loss term in the interaction
Hamiltonian. Considering only atom-atom collisions involving one atom
with wavevector $\kko$ and one with $\kkt$, Eqs.~(\ref{opeq1}) and
(\ref{opeq2}) in the classical field approximation become
\begin{equation}
  \begin{split}
    \label{opeql}
    i \derfrac{}{\tau} \tilde{a}_1 &=
    3\,\tilde{a}_1^*\,
    \tilde{a}_1^*\,
    \tilde{b}_1\,
    \tilde{a}_2\,
    +\,
    \tilde{b}_2^*\,
    \tilde{a}_2\,
    \tilde{a}_2\,
    \tilde{a}_2\,
    - i \gamma \,|\tilde{a}_2|^2 \, \tilde{a}_1
    \,, \\
    i \derfrac{}{\tau} \tilde{a}_2 &=
    3\,\tilde{a}_2^*\,
    \tilde{a}_2^*\,
    \tilde{b}_2\,
    \tilde{a}_1\,
    +\,
    \tilde{b}_1^*\,
    \tilde{a}_1\,
    \tilde{a}_1\,
    \tilde{a}_1\,
    - i \gamma \,|\tilde{a}_1|^2 \, \tilde{a}_2
    \,.
  \end{split}
\end{equation}
Here the parameter $\gamma$ is the elastic scattering rate constant
divided by 2$\hbar \Omega$, and is given by $\gamma=(\hbar k_0/m)\,
\sigma n /(2 \hbar \Omega)$~\cite{band2000}, where $\sigma$ is the
elastic scattering cross section at the relevant two-body collision
energy $E_{\text{coll}} =\hbar^2 k_0^2 /m $.  Eqs.~(\ref{opeql}) show
that the two processes --- resonant three-body recombination and
condensate loss due to elastic scattering --- compete with each other
and that the value of the parameter $\gamma$ is crucial in determining
which process dominates.  Fig.~\ref{fighes} shows the solution of the
Heisenberg equations that include the elastic loss terms for two
extreme cases. The example with $\gamma=0.1$ illustrates the case of
little elastic loss allowing almost $70$\% of the atoms to be
transformed into molecules if $t_{\text{coll}}\approx \TR$. However,
if $\gamma=1.5$, for the same collision time more than $80$\% of the
atoms are lost due to elastic collisions and only about 8\% end up in
the molecular clouds.

The simplest estimation of $\gamma$ can be attempted in the low energy
limit ($k_0\ll a^{-1}$) where $\sigma \approx 8 \pi a^2$.  This leads
to $\gamma= (8 \pi/\sqrt{3} \zeta) (n a ^3)^{-1/2}$ and yields high
values of $\gamma$ even for very large values of the diluteness
parameter (i.e., $n a ^3 =1$ corresponds to $\gamma=1.5$).  However,
note the importance of using an exact value of $\sigma$ corresponding
to the correct $E_{\text{coll}}$.  For example, in the case of a
weakly bound molecular level [$E_0\approx \hbar^2 /(m a^2)$],
$k_0\approx1/a$, which is beyond the validity of the low energy
approximation.  Apparently, changing the value of the parameter $n
a^3$ by changing either $n$ or $a$ is the primary way of shifting the
balance between the two processes.  However, special conditions might
enhance or suppress either process.  One such situation may occur when
$a$ approaches zero and therefore elastic scattering vanishes while
three-body recombination still has a considerable probability.
Calculations by Esry {\it et al.}~\cite{esry99} predict a significant
$K_3$ for $a=0$.  Similarly, Stenger {\it et al.}~\cite{stenger99}
measured three-body recombination loss from Na condensates in the
vicinity of Feshbach resonances, observing a large amount of loss even
at small values of $a$.  Consider the Feshbach resonance located near
$B=535.7\,$G in $^{85}$Rb.  Tuning the magnetic field such that $a=0$
($B=537.6$\,G) leads to $E_0=75\,\mu$K, and for a density of
$10^{16}/{\rm cm}^3$ we obtain $K_3=5.4 \times 10^{-32}$\,cm$^6$/s (as
estimated in Ref.~\cite{esry99}), giving $\TR\approx 28$\,ms.  For Na
with $a=0$ ($B=907$\,G), Ref.~\cite{stenger99} measures
$K_3\approx10^{-28}$\,cm$^6$/s, giving $\TR\approx 0.24$\,ms.
Alternatively, one can choose the experimental conditions such that
the elastic collisions are strongly suppressed exactly at the resonant
value of the collision energy $E_{\text{coll}}$ due to the presence of
a Ramsauer minimum (see, for example,~\cite{Burke98}).  The exact
resonant matching between the position of the Ramsauer minimum and
that of the binding energy of the molecular state (i.e., $E_{\rm
coll}=E_0/3$) can be achieved near some Feshbach resonances. For the
above mentioned resonance in $^{85}$Rb, this matching occurs for
$B=547$\,G, when $a=-353$\,a.u., and $E_0=1.05$\,mK.  In this case,
considering $n=10^{15}/{\rm cm}^3$, one finds $\TR\approx 1.3$\,ms
($K_3=9.2 \times 10^{-26}$cm$^6$/s cf.~\cite{esry99}). Since producing
an atomic condensate at negative $a$ might be a problem, we assume
here that it is produced at some positive $a$ and that the magnetic
field is swept into resonance precisely when the two atomic clouds
start colliding.

In conclusion, resonant three-body recombination that benefits from
the superradiant-like bosonic stimulation could be a promising
candidate for producing coherent samples of molecules from existing
atomic condensates.  The simplest experimental setup that takes
advantage of this process involves a collision of two atomic
condensates, which produces two molecular condensates in
counterpropagating momentum eigenstates.  This setup could be
particularly promising for the construction of a molecule laser by
colliding two atomic condensates.

This work was supported in part by the NSF.

\end{document}